\documentclass{PoS}

\usepackage{aas_macros} 
\usepackage{url} 

\newcommand{\nggn}{$(n,\gamma)\rightleftarrows(\gamma,n)$}  

\title{Nuclear masses near $N=82$ that influence $r$-process abundances}
\ShortTitle{Nuclear masses near $N=82$ and $r$-process abundances}

\author{\speaker{Matthew Mumpower}%
        \thanks{We thank Grant Mathews for helpful discussions. This work was supported by the NSF through the Joint Institute for Nuclear Astrophysics grant numbers PHY0822648 and PHY1419765. }\\
        University of Notre Dame\\
        E-mail: \email{matt.mumpower@nd.edu}}

\author{Rebecca Surman\\
        University of Notre Dame\\
        E-mail: \email{rsurman@nd.edu}}

\author{Mary Beard\\
        University of Notre Dame\\
        E-mail: \email{mbeard@nd.edu}}
        
\author{Dong-Liang Fang\\
        Michigan State University\\
        E-mail: \email{Fang@nscl.msu.edu}}

\author{Ani Aprahamian\\
        University of Notre Dame\\
        E-mail: \email{aapraham@nd.edu}}

\abstract{Nuclear masses are one of the key ingredients of nuclear physics that go into astrophysical simulations of the $r$ process. 
Nuclear masses effect $r$-process abundances by entering into calculations of Q-values, neutron capture rates, photo-dissociation rates, beta-decay rates, branching ratios and the properties of fission. 
Most of the thousands of short-lived neutron-rich nuclei which are believed to participate in the $r$ process lack any experimental verification, thus the identification of the most influential nuclei is of paramount importance. 
We have conducted mass sensitivity studies near the $N=82$ closed shell in the context of a main $r$-process. 
Our studies take into account how an uncertainty in a single nuclear mass propagates to influence the relevant quantities of neighboring nuclei and finally to $r$-process abundances. 
We identify influential nuclei in various astrophysical conditions using the FRDM mass model. 
We show that our conclusions regarding these key nuclei are still retained when a superposition of astrophysical trajectories is considered. 
}

\FullConference{XIII Nuclei in the Cosmos\\
                 7-11 July, 2014\\
                 Debrecen, Hungary}

\begin{document}

\section{Introduction}
One of the most challenging open questions driving the focus of research in nuclear astrophysics is the astrophysical location of the rapid neutron capture process or $r$ process of nucleosynthesis \cite{BBFH1957}. 
This complex problem ties together theoretical models of nuclear physics, experimental measurements of short-lived nuclei and astrophysical observations. 
Simulations of the $r$ process are well constrained by observations of very old stars as they are believed to contain the enrichment of only a few events \cite{Sneden2008}. 
Observations of these `metal-poor' stars is limited to elemental abundances, and thus solar $r$-process residuals are also studied which provides additional isotopic information. 
The limitation of the solar residuals is that it a convolution of many events that occurred before the formation of the solar system \cite{Arlandini1999}. 
To attempt to match the features found in these observed abundances, studies of the $r$ process have probed a range of astrophysical environments \cite{Arnould2007}. 
When experimental data is unavailable, simulations of the $r$-process use models of nuclear physics to extrapolate the properties of unmeasured nuclei such as masses, rates and branching ratios, all of which greatly impact the predictions of $r$-process abundances. 
Our group has recently surveyed how individual nuclear properties influence predictions of $r$-process abundances by exploring variations in individual neutron capture rates \cite{Mumpower2012c,AIP-NC}, $\beta$-decay rates \cite{AIP-Beta}, and binding energies \cite{Brett2012,AIP-BE}. 
This series of papers has culminated in the study of how uncertainties in nuclear masses propagate self-consistently to these quantities and ultimately alter $r$-process predictions \cite{Mumpower2014a}. 
We have since generalized this work for a range of separate astrophysical conditions \cite{Mumpower2014b}. 

In this contribution we lay the groundwork for expanding our self-consistent mass sensitivity studies using an abundance pattern that more closely approximates the solar $r$-process residuals. 
By considering a superposition of previously studied trajectories from Ref. \cite{Mumpower2014b} we probe the question: \textit{Do our previous predictions of influential nuclear masses remain when astrophysical conditions are combined?} 
The details of the astrophysical conditions, network calculations and separate sensitivity studies explored here are covered in Ref. \cite{Mumpower2014b}. 
We note that FRDM masses are used as the basis for all rate and branching ratios calculations \cite{FRDM1995}. 

\section{Propagation of uncertainties}
We propagate nuclear mass variations to all the relevant nuclear properties of neighboring nuclei that depend on the changed mass. 
When we vary the mass of a nucleus ($Z$,$A$) with $Z$ protons, $N$ neutrons and $A=Z+N$ total nucleons we follow the changes to the neutron capture rates of ($Z$,$A$) and ($Z$,$A-1$), the separation energies of ($Z$,$A$) and ($Z$,$A+1$), the $\beta$-decay rates of ($Z$,$A$) and ($Z-1$,$A$), and $\beta$-delayed neutron emission probabilities of ($Z$,$A$), ($Z-1$,$A$), ($Z-1$,$A+1$), ($Z-1$,$A+2$), ($Z-1$,$A+3$) and ($Z-1$,$A+4$). 

Statistical model calculations require a number of nuclear inputs. 
The present calculations have been performed using the publicly available statistical model code TALYS \cite{TALYS}, with default settings for level density, $\gamma$-strength function and particle optical model. 
Nuclear masses can enter into the neutron capture rate calculations through the choice of these models \cite{Beard2014}. 
The default level density model used in TALYS combines a constant nuclear temperature at low energies, matched with the back-shifted Fermi gas model at some matching point energy which is determined via systematics \cite{CT}. 
Masses can enter into the former of these two via the definition of the constant temperature, which, as used in the TALYS code, is proportional to one over the square root of the shell correction term: $dW(Z,A) = M(Z,A)-M_{LDM}(Z,A)$ where $M(Z,A)$ is the mass of the nucleus (experimental if available) and $M_{LDM}(Z,A)$ is the predicted mass to a spherical liquid-drop. 
Additionally, the level density parameter $a$, used in the back-shifted Fermi gas term of the level density, also depends on $dW$ and hence the nuclear masses. 
In this case, $a$ is proportional to $dW$. 
For the $\gamma$-strength function, the default setting in TALYS is to use the formulation set out in Kopecky-Uhl (KU) \cite{KU}. 
Nuclear masses enter into this parameterization of the giant dipole resonance again via a nuclear temperature term which prevents the $\gamma$ strength from going to zero as $\gamma$ energy decreases. 
The nuclear temperature itself is proportional to the square root of the neutron separation energy, $S_n$, as well as $a$, evaluated at $S_n$. 

Photo-dissociation rates are calculated from neutron capture rates by detailed balance: 
\begin{equation}\label{eqn:photo}
\lambda_\gamma(Z,A) \propto T^{3/2} \exp\left[-{\frac{S_n(Z,A)}{kT}}\right] \langle \sigma v \rangle_{(Z,A-1)}
\end{equation}
where $S_n(Z,A)= M(Z,A-1)-M(Z,A)+M_n$ is the one neutron separation energy, $M(Z,A)$ is the mass of the nucleus with $Z$ protons and $A$ nucleons, $M(Z,A-1)$ is the mass of the nucleus with one less neutron, $M_n$ is the mass of the neutron, $T$ is the temperature, $\langle \sigma v \rangle_{(Z,A-1)}$ is the neutron capture rate of the neighboring nucleus and $k$ is Boltzmann's constant. 
From Eqn. \ref{eqn:photo} we see that if $M(Z,A)$ is altered it can exponentially influence the rate, leading to large changes in final abundances \cite{Surman2009}. 

We use the Quasi-Particle Random Phase (QRPA) model from Ref. \cite{DLF2013} to calculate weak decay properties near the $N=82$ closed shell. 
This model includes both Gamow-Teller (GT) and First-Forbidden (FF) transitions along with realistic forces. 
To recalculate the $\beta$-decay rate, $\lambda_\beta$, when a mass is varied we compute: 
\begin{equation}\label{eqn:beta}
\lambda_\beta=\sum_{0\leq E_i\leq Q_{\beta}} f(Q_{\beta}-E_i) S_{\beta}(E_i)
\end{equation}
where $i$ denotes the i\textsuperscript{th} excited state of the daughter nucleus with energy 
$E_i$, $f$ is the phase space factor and $S_{\beta}$ is the $\beta$-strength function obtained from the QRPA solutions. 
We have explored the dependence of masses on $S_{\beta}$ and found it to be very small. 
Therefore, when a mass is varied we only recalculate the phase space factor. 
For allowed transitions, $f$ can be approximated by $f\sim(Q_{\beta}-E_i)^5=(M(Z,A)-M(Z+1,A)-E_i)^5$ which shows the general dependence of the half-life on masses. 

In our propagation of mass variation to weak decay properties we assume that the structure of the ground state does not change. 
Therefore, a modification in predicted neutron branching ratios comes from a change in the $\beta$-decay Q-value of the parent nucleus, $Q_{\beta}$, or from a change to the neutron separation energies, $S_n(Z,A)$, of the daughter nuclei. 
This can be seen from the expression for neutron emission probabilities: 
\begin{equation}\label{eqn:pn}
P_n=(\sum_{E_i<S_n}\lambda_i)/(\sum_{E_i<Q_{\beta}}\lambda_i)
\end{equation}
where $E_i$ is the excitation energy of the i\textsuperscript{th} state and $\lambda_i$ is the decay-width to the i\textsuperscript{th} state. 
Since a variation in the mass, $M(Z,A)$, changes both $Q_{\beta}$ and $S_n$ the general impact is difficult to generalize. 

\begin{figure*}
\centerline{\includegraphics[width=\textwidth]{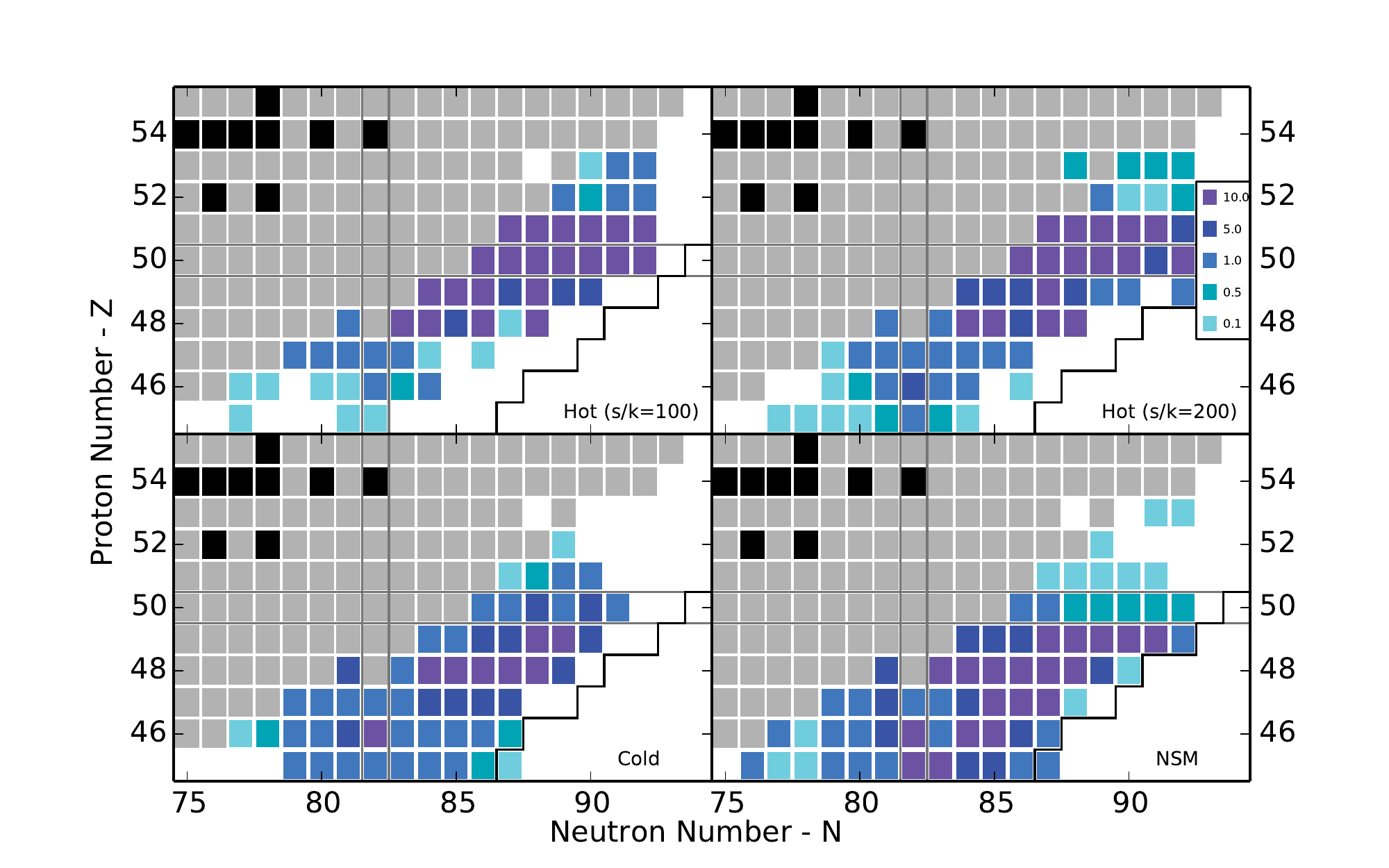}}
\caption{Maximum sensitivity measures $F_{max}(Z,A)$ for FRDM nuclear masses (variation by $\pm1.0$ MeV) near the $N=82$ closed shell for four $r$-process trajectories from \cite{Mumpower2014b}. 
Black squares are stable isotopes and gray shading indicates the region of measured nuclear masses from the 2012 AME. 
An estimated accessibility limit for FRIB is shown as a black line \cite{FRIB2014}.}
\label{fig-1}
\end{figure*}

\section{Results}
In our sensitivity studies we quantify the effect masses have on final abundances by computing the metric, $F$. 
This quantity is defined as the difference between the final composition in the unchanged, baseline simulation and the final composition in the simulation where a single nuclear mass was varied \cite{Mumpower2014b}. 
We contrast the results of nuclear mass sensitivity studies for four trajectories in Fig. \ref{fig-1}. 
These studies were performed with vastly different astrophysical conditions: a hot wind with entropy $s/k=100$, a second hot wind with entropy $s/k=200$, a cold wind and neutron star merger. 
Nevertheless, we find that a predictable trend in the distribution of influential nuclei emerges: 
In both hot $r$-process simulations (top two panels) where the path, or set of most abundant isotopes, is initially set by \nggn \ equilibrium we find a shift in the distribution of influential nuclei to higher $Z$ and $N$. 
This is caused by the long duration equilibrium phase which prevents the path from extending to the neutron dripline. 
Fission recycling is active in the cold wind and the neutron star merger trajectory which weights the $r$-process path with material in this region at lower $A$ closer to the closed $N=82$ shell, resulting in sensitivities to masses at lower $A$. 

\begin{figure*}
\centerline{\includegraphics[width=19cm]{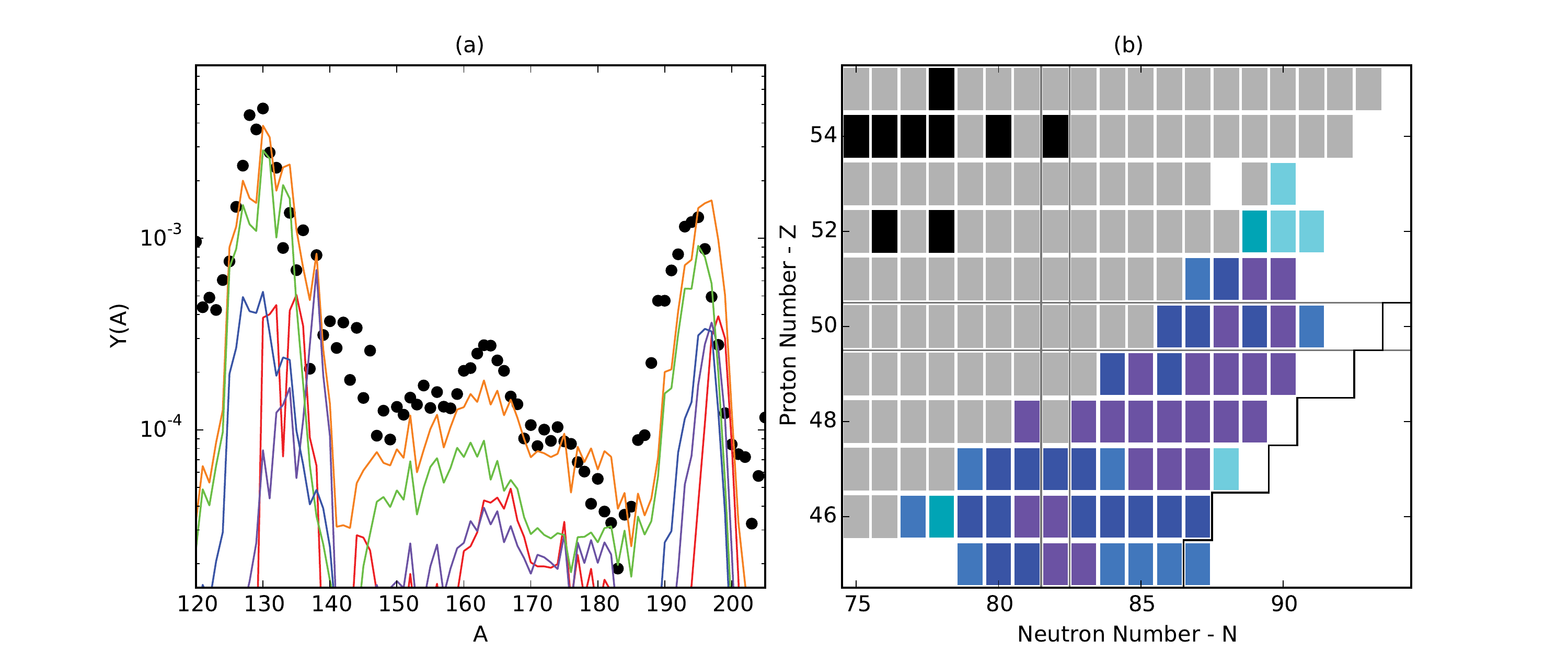}}
\caption{(a) Baseline abundance pattern (orange) from a superposition of four astrophysical trajectories: hot $s/k=100$ (red), hot $s/k=200$ (purple), cold (blue) and neutron star merger (green). Black dots are solar $r$-process residuals from Ref. \cite{Arlandini1999}. (b) New sensitivity measure $F^{new}_{max}(Z,A)$ for FRDM nuclear masses (variation by $\pm1.0$ MeV) when using a superposition of the four trajectories.}
\label{fig-2}
\end{figure*}

We now address the question of whether the distribution of influential nuclei transforms when a superposition of these trajectories is considered. 
To approach this question we combine the abundances from the separate trajectories using a weighted sum of the final patterns as shown in panel (a) of Fig. \ref{fig-2}. 
The four trajectories: hot wind $s/k=100$ (red), hot wind $s/k=200$ (purple), cold wind (blue) and neutron star merger (green) have weights $w_i=0.8$, $1.0$, $1.5$ and $3.5$ respectively which lead to a final baseline pattern (orange). 
This same weighting is then used to compute the new final abundances for all nuclei in our sensitivity studies. 
For each nuclei in our study there are two final patterns: one where the mass was increased by $1.0$ MeV and a second where the mass of the nucleus was decreased by $1.0$ MeV. 
It is important to note that this new set of abundance patterns may produce results that do not resemble the conclusions that were drawn from the investigation of the separate trajectories. 
Due to the superposition some final abundance signatures may cancel out and others may be reinforced. 
To probe the extent of this effect we recalculate the maximum $F$-value, $F^{new}_{max}(Z,A)$, between the $\pm1.0$ MeV cases with the new superposition of abundances, see panel (b) of Fig. \ref{fig-2}. 
Remarkably, we find that the same overall distribution of influential nuclei persists through the superposition: the hot $r$-process trajectories contribute to $F^{new}_{max}(Z,A)$ near higher $A$ while the cold and neutron star merger trajectories highlight sensitivities at lower $A$. 

\section{Conclusions}
We have performed self-consistent mass sensitivity studies for nuclei near the $N=82$ closed neutron shell with a variety of astrophysical conditions. 
We investigated a superposition of these trajectories and asked the question: 
Do our previous predictions of influential nuclear masses remain when astrophysical conditions are combined? 
\textit{In these preliminary calculations we have shown our conclusions drawn regarding key nuclei from separate mass sensitivity studies remain intact even when a superposition of vastly different astrophysical trajectories is considered}. 
This shows the robustness of our results for these key nuclei and also provides further motivation for the measurements of neutron-rich nuclei that are within reach of future radioactive beam facilities. 

\bibliographystyle{unsrt}
\bibliography{refs}

\end{document}